\documentclass[a4paper,aps,pra,showpacs,preprintnumbers,twocolumn]{revtex4}
\usepackage{graphics,graphicx,epsfig,epstopdf}
\usepackage[centertags]{amsmath}
\usepackage{amsmath,amsfonts}
\usepackage{amssymb}
\usepackage[cp1251]{inputenc}
\usepackage[english]{babel}
\inputencoding{cp1251}
\usepackage{graphicx}
\usepackage{amsthm,color}

\begin{document}

\renewcommand{\Im}{\mathop{\mathrm{Im}}\nolimits}
\renewcommand{\Re}{\mathop{\mathrm{Re}}\nolimits}

\title{Quantum computations on the ensemble of two-node cluster states, obtained by sub-Poissonian lasers}
\author{S. B.\, Korolev }
\email{Sergey.Koroleev@gmail.com}
\author{A. N.\, Dobrotvorskaia}
\author{T. Yu.\, Golubeva}
\author{Yu. M.\, Golubev}
\affiliation{St. Petersburg State University, Universitetskaya  nab. 7/9, 199034 St. Petersburg, Russia}
\date{\today}
\begin{abstract}
In this study, we demonstrate the possibility of the implementation of universal Gaussian computation on a two-node cluster state ensemble. We consider the phase-locked sub-Poissonian lasers, which radiate the bright light with squeezed quadrature, as the resource to generate these states.
\end{abstract}
\maketitle
\section{Introduction}
One of the fundamental problems in modern science is the creation of a quantum computer. The most compelling feature of such a device, as compared to conventional computers, is the theoretical capability to solve exponential problems (factorization of large numbers, quantum system modelling, etc.) effectively by virtue of inherent quantum parallelism and using of a superdense coding protocol.

Quantum computer works by taking quantum states as input and transforming them into output states. This operation is universal in the sense that it can execute any kind of unitary transformation. There are a number of different realization of quantum computing: circuit, adiabatic, topological, etc. Quantum computers can also be classified by the type of quantum systems which they are based upon: characterized by either discrete or continuous variables.

This article is concerned with one-way quantum computations with continuous variables \cite{Menicucci_1}. This kind of computation involves the local measurements of a physical system in a quantum cluster state \cite{Raus2}. Each element of such a system interacts with one or several other elements due to the quantum entanglement, forming a complex physical structure. This structure can be visualized by a graph where each element of the system is represented by a node and each quantum entanglement by an edge. According to the theorem of reduction, the measurement of a quantum system unavoidably disrupts it, though at the same time modifying subsystems connected to it via entanglement in a predictable way. Type of transformation that can be obtained by calculations in a quantum computer is determined by the topology of the corresponding graph. More complex and rich topologies are characterized by a larger number of different kinds of transformations that can be realized.

When dealing with continuous variables a particular class of physical systems is used, which serve as a resource for one-way computations. These states are generated with the help of squeezed quantum oscillators. Early works concerned with proving the conceptual possibility of quantum calculations based on such systems avoided discussions of the squeezing degree, which was presumed to be infinitely large. It is obvious, though, that in order to determine the computational capabilities of real systems the limitations to quantum squeezing must be considered. As shown in \cite{Korolev_2018} the topology of cluster state depends directly on squeezing degree of quantum oscillators, involved in the generation of this state: the lower the squeezing, the smaller the possible number of nodes and edges of the cluster state.

As we mentioned before, an ideal quantum computer should be able to perform any kind of transformation. For quantum computations with continuous variables, this universality manifests in capability to realize two primal kinds of operations: universal multimode Gaussian and single-mode non-Gaussian \cite{Lloyd}. By definition, Gaussian quantum transformation modifies an input state in linear way, while non-Gaussian transformation is non-linear. Thus, if a particular system can be demonstrated to be capable of accomplishing both types of transformations, it can be concluded that any logical operation can be processed by this system.

The main idea and originality of this study consists in the performing the universal multimode Gaussian transformations on an ensemble of two-node cluster states (states with the simplest topology), without the use of clusters with a richer structure. This approach has undoubted advantages (described in detail in Section II of the manuscript). First of all, it is an opportunity to carry out computations, having as a resource the squeezed light with a lower degree of squeezing. In fact, the larger the number of edges between nodes in the cluster state, the more squeezed light is required to generate it \cite{Korolev_2018}. In Ref. \cite{Menicucci_2011} the authors employed cluster states with a more complex topology to implement universal Gaussian transformations. In our case, we carry out all computations directly on two-node cluster states that allow us to use the minimally squeezed light. In addition, our scheme has other simplifications: to create two-node cluster states, only two sources of squeezed light are needed, instead of four in Ref. \cite{Menicucci_2011}. Finally, a much smaller number of linear optical elements is sufficient to realize this scheme, which is essential for experimental implementation.  In view of the above, we believe that such an issue is not only theoretical but also practical interest. This work does not deal with the realization of single-mode non-Gaussian transformation since this problem has been already theoretically implemented for the two-node cluster state \cite{Gu}. There is also a series of other theoretical proposals to implement a single-mode non-Gaussian transformation \cite{Marek, Yukawa, Park, Arzani}, which lie far from the main subject of this article.

There are several approaches to arranging required two-node cluster states. In this work, the radiation of phase-locked sub-Poissonian laser is used as a source of squeezing for generation of two-node cluster states ensemble. Two features of such lasers are crucial for quadrature squeezing of light: regularity of pumping and addition of weak external synchronizing signal. The regularity of pumping suppresses quantum fluctuation in the number of photons in laser’s output radiation (sub-Poissonian photon statistics). On the other hand, the addition of external field restrains phase diffusion and locks the phase. The main advantage of these emitting sources is that they can be easily manufactured from semiconducting materials \cite{Kim}. It appears that for such materials the requirements for regular pumping are satisfied naturally without additional efforts. The additional advantage of phase-locked sub-Poissonian laser is the intensity of produced light and its characteristic space-time profile, which enables the use of this light for homodyne detection of a signal after all calculations. This property is lacking in an optical parametric amplifier, for example, where implementing homodyne with required emission profile is compulsory. Among the disadvantages of the sub-Poissonian lasers relatively small squeezing degree of emitted light quadrature can be mentioned  \cite{Yamamoto}. Nonetheless, these levels are absolutely sufficient for generation of two-node cluster states.

This paper is organized as follows. The second section defines cluster states for continuous variables and discusses limitations imposed by squeezing degree. In the third section generation of two-node cluster states are described as realized in the phase-locked sub-Poissonian laser. Section IV illustrates the procedures of single-mode and two-mode computations of input signals by the ensemble of two-node cluster states. The appendix is provided with detailed calculations verifying that van Loock-Furusawa criterion is satisfied.

\section{Minimum number of nodes in the cluster state}
Cluster state is defined as a multipartite entangled state characterized by a mathematical graph G: each physical subsystem constituting the cluster state is represented by a node and each quantum entanglement between subsystems forms an edge.

This study deals with continuous variable cluster states only. The description of such states is based on a concept of quantum oscillator, i.e.,  a quantum-mechanical system described by quadratures $ \hat{x} $ and $ \hat{y} $, which obey the canonical
commutation relation:
\begin{align}
\left[\hat{x},\hat{y}\right]=\frac{i}{2}.
\end{align}
The key requirement for generation of the cluster states is that all engaged oscillators have to be squeezed. For consistency of the construction scheme, it is assumed that the $\hat{y}$-quadratures are squeezed and the following inequality holds for the corresponding variances:
\begin{align}
\langle \delta \hat{y}^2 \rangle \textless \frac{1}{4},
\end{align}

The next step after the squeezed state of the oscillators is obtained is to entangle them. In this regard, the unitary transformation of the general form \cite{Korolev_2018, HAL} has to be performed:
\begin{align} \label{1}
U=\left(I+iA \right)\left(I+A^2\right)^{-1/2}Q.
\end{align}
Here $A$ is the adjacency matrix of the graph $G$, which defines the cluster topology, and $Q$ is an arbitrary orthogonal matrix. This operation transforms a vector consisting of squeezed quadratures of oscillators $\hat{x}_j+i\hat{y}_j$ into the one formed by cluster state quadratures
$\hat{X}_j+i\hat{Y}_j$:
\begin{align}
\begin{pmatrix}
\hat{X}_1+i\hat{Y}_1\\
\vdots \\
\hat{X}_n+i\hat{Y}_n
\end{pmatrix}=U\begin{pmatrix}
\hat{x}_1+i\hat{y}_1\\
\vdots \\
\hat{x}_n+i\hat{y}_n
\end{pmatrix}.
\end{align}
The obtained cluster state has specific topology defined by the adjacency matrix $A$. This transformation ensures that the variances of nullifier operators $\hat{N}_j$ tend to zero in the limit of infinite squeezing of all quantum harmonic oscillators used to generate the cluster state \cite{Korolev_2018}:
\begin{multline} \label{2}
\forall j \in \lbrace 1,2, \dots ,n\rbrace \quad  \lim \langle \delta \hat{N}_j ^2\rangle=0 \\
 \text{for} \quad \langle \delta \hat{y}_1 \rangle
\rightarrow 0, \langle \delta \hat{y}_2^2 \rangle \rightarrow 0, \dots , \langle \delta \hat{y}_n^2 \rangle \rightarrow 0.
\end{multline}
Recall that nullifiers $\hat{N}_j$ are operators which represent the linear combinations of canonical quadratures of a cluster state,
\begin{align} \label{3}
\hat{N}_j=\hat{Y}_j-\sum _{i=1}^n\left[A\right]_{ji}\hat{X}_i ,\qquad j=1, \dots , n,
\end{align}
where $\left[ A\right]_{ji}$ are elements of the adjacency matrix, $n$ is the number of nodes in the graph. The relationship (\ref{2}) is generally thought as the definition of the cluster state \cite{vLoock}. This definition refers to the concept of the ideal cluster state (with zero nullifiers variances), which requires infinitely squeezed oscillators for the generation. Obviously, the infinite squeezing is impossible for the real physical systems, which restricts the possibility of cluster generation.

The limitation to the degree of oscillators’ squeezing affects primarily the topology of cluster states generated on the base of these oscillators \cite{Korolev_2018}. For the case when $\hat{y}$-quadratures of all oscillators are squeezed equally, the number of nodes in cluster state and variances $\langle\delta y^2 \rangle$ of each $\hat{y}$-quadratures are related as follows:
\begin{align} \label{4}
\langle\delta y^2 \rangle < \min_{(i,j)} \left[\frac{ 1}{2+\deg(i)+\deg (j) } \right].
\end{align}
Here the minimum on the right-hand side is taken over all the pairs of the adjacent nodes $i$ and $j$, $\deg (i)$ and $\deg (j)$ are the degrees of the nodes $ i $ and $ j $ (the number of neighbors of nodes $ i $ and $ j $), respectively.  As this relationship implies, in order to create a cluster state with branching structure the squeezing degree of oscillators used for generation of this state has to be increased. It was demonstrated \cite{Menicucci} that a cluster state produced from oscillators squeezed by 20 dB can be used to perform universal multimode Gaussian transformations with error correction. However, to date, the maximum experimentally implemented squeezing degree is 15 dB \cite{Vahlbruch}, so the universal computer, working on the proposed algorithm \cite{Menicucci} is not realizable today.

Inequality (\ref{4}) dictates to us the desire to simplify the cluster structure as much as possible, which would reduce the requirement for the squeezing degree of oscillators for its construction. At the same time, we should be able to perform a universal set of logical operations by this state. Taking this into account, the presented study aims at the realization of universal multimode Gaussian transformation not on a single multi-node cluster state, but on an ensemble of cluster states with a smaller number of nodes. As an example of such ensemble, an ensemble of two-node cluster states will be considered. The number of nodes in such state is minimal, therefore their generation requires oscillators with the least squeezing. It follows from inequality (7) that such squeezing should exceed one fourth, i. e., oscillators should be at least slightly squeezed.

\section{Generation of two-node cluster state ensemble by the phase-locked sub-Poissonian laser}
In this section, the generation of cluster states composed of two-nodes is considered. Speaking about the generation, first of all, it is necessary to choose the sources by which this generation will be carried out. In the case under study two phase-locked sub-Poissonian lasers are taken as these sources, which emit the stationary light in squeezed state \cite{Golubeva}. Since two-node cluster states are generated, there is no need to impose special conditions on the squeezing degree of radiation.

\begin{figure*}
\centering
\includegraphics[scale=1]{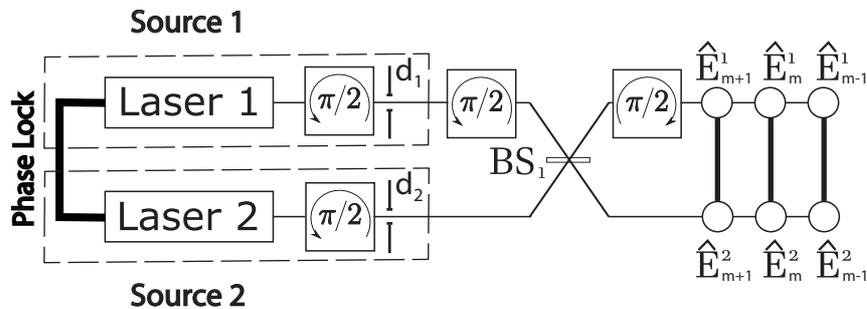}
\caption{Scheme of the generation of two-node cluster state sequence by two synchronized sub-Poissonian lasers. Here $\mbox{BS}_1$ is a beam splitter, $\mbox{d}_j$  is an aperture in $j$'th channel.} \label{Fig_1}
\end{figure*}

In order to generate clusters and then carry out measurements, the stationary light beam should be converted to pulses. For this purpose, two apertures can be placed near each laser which is able to shut and open periodically (see. Fig. \ref{Fig_1}). Apertures stay opened for time period $T$ and shut for time period  $T_0$. If open state intervals of both apertures are correlated, then two light flows will be produced which consist of pulses with duration $T$ and separated from each other in time by $T_0$. Every flow of pulses is indexed by $j=1,2$ and matched with Heisenberg amplitudes
\begin{align}
\hat{S}_{j}(t) = \sum _{m=1}^{n} \hat{S}^j_{m}(t)\;\Theta(t-t_m)+\hat V_{j}(t),\qquad j=1,2 .
\end{align}
Here $t_m=(m-1)(T+T_0)$ is the start time of the m'th light pulse, $\Theta(t-t_m)$ is the window function equal to one at $t_m<t<t_m+T$ and to zero outside of these time intervals. The value $\hat{S}^j_{m}(t)=\hat{x}^j_m(t)+i\hat{y}_m^j(t)$  corresponds to the Heisenberg amplitude of $m$'th pulse in the $j$'th light beam, operators $\hat{x}_m^j(t)$ and $\hat{y}_m^j(t)$ are quadratures of this pulse. The term $\hat V_j(t)$ determines the contribution of vacuum channels and its presence in the equation allows matching commutation relations with each other
\begin{align}
&\left[\hat{S}_{j}(t),\hat{S}^\dag_{j'}(t^\prime)\right] =\delta_{jj'}\delta (t-t^\prime),\\
&\left[\hat{S}_{m}^j(t),\left(\hat{S}^{j'}_{m'}(t^\prime)\right)^\dag\right] =\delta_{jj'}\delta_{mm'}\delta (t-t^\prime).
\end{align}
It was demonstrated \cite{Samburskaya} that making the duration of each of pulses much longer than a length of laser correlations, guarantees that the pulses will be well squeezed, exactly like original stationary laser radiation. Let us note, that while a phase-locked sub-Poissonian laser emits light squeezed by $\hat{x}$-quadrature, it can be converted to $\hat{y}$-quadrature by rotating radiation phase by ${\pi}/{2}$. Taking into account discussed above, the following equation for $\hat{y}$-quadratures of produced light pulses can be formulated:
\begin{multline} \label{5}
\langle: \delta\hat{y}^{j}_{m}(t)\;\delta\hat{y}^{j^\prime}_{m'}(t^\prime):\rangle =-\frac{\kappa}{8}\frac{1-\mu}{1-\mu/2}\; e^{\displaystyle-\kappa
(1-\mu /2)|t-t^\prime| }  \\
 \times \Theta(t-t_m)\Theta(t^\prime-t_{m'})\delta_{mm'}\delta_{jj^\prime}.
\end{multline}
Here $\langle : \;: \rangle$ means the average of normally ordered operators, parameter $\kappa\gg T^{-1} $ is the spectral width of laser resonator mode. Parameter $\mu$ characterize the measure of synchronization of laser generation. It is assumed that laser generation is affected by synchronizing fields very slightly, i.e. $\mu\ll1$. By using this approximation in the equation (\ref{5}) the following simple equality is obtained:
\begin{multline}
\langle: \delta\hat{y}^{j}_{m}(t)\;\delta\hat{y}^{j^\prime}_{m'}(t^\prime):\rangle =-\frac{\kappa}{8}\; e^{\displaystyle-\kappa |t-t^\prime| }  \\
\times\Theta(t-t_m)\Theta(t^\prime-t_{m'})\delta_{mm'}\delta_{jj^\prime}.
\end{multline}

Once the sources of the squeezed radiation were determined, we can proceed to the scheme for generating a two-node cluster state. For this purpose the unitary transformation should be defined, that converts the light pulses from the squeezed state to the cluster one. In order to obtain the matrix of such transformation the Eq. (\ref{1}) can be utilized. Let us substitute in this equation the adjacency matrix for the two-node graph:
\begin{align}
A^{(2)}=\begin{pmatrix}
0 & 1 \\
1 & 0
\end{pmatrix}.
\end{align}
As a result, the unitary transformation matrix takes the form:
\begin{align} \label{6}
U=\frac{1}{\sqrt{2}}\begin{pmatrix}
1 & i\\
i & 1
\end{pmatrix}Q.
\end{align}
Here there is an arbitrary orthogonal matrix $Q$. It should be emphasized that the particular form of the matrix $Q$ does not affect the topology of the cluster state and the values of nullifiers \cite {Korolev_2018}, although this matrix defines a unitary transformation $U$, and hence the procedure for creating a cluster state. Accordingly, any orthogonal matrix can be chosen which simplifies calculations. If as such a matrix one choose
\begin{align}
Q=\begin{pmatrix}
1 & 0 \\
0 & -1
\end{pmatrix},
\end{align}
then the unitary matrix (\ref{6}) can be split into a product of three matrices of the form
\begin{align}\label{expand}
U=\frac{1}{\sqrt{2}}\begin{pmatrix}
1 & -i\\
i & -1
\end{pmatrix}=\begin{pmatrix}
-i & 0\\
0 & 1
\end{pmatrix} \;  \begin{pmatrix}
\frac{1}{\sqrt{2}} & \frac{1}{\sqrt{2}}\\
\frac{1}{\sqrt{2}} & -\frac{1}{\sqrt{2}}
\end{pmatrix} \; \begin{pmatrix}
i & 0\\
0 & 1
\end{pmatrix}.
\end{align}
Scheme of the experiment corresponding to the specified transformation is demonstrated in Fig.\ref{Fig_1}. First and third matrices of decomposition (\ref{expand}) rotate laser pulse phases in the upper arm by the angles $\pi/2$ and $-\pi/2$, respectively. The second matrix is the matrix of beam splitter transformation. Therefore, in order to create a two-node cluster state, it is necessary to apply a sequence of three transformations to squeezed light pulses: firstly, to rotate the phase of the first pulse by the angle $\pi/2$, next to transform on the beam splitter, and finally to rotate the phase of the first pulse by the angle $-\pi/2$.
After passing through this scheme initially squeezed pulses are to become the cluster state pulses. In order to be sure that this indeed happens, the nullifiers of the resulting state need to be verified. Since the real sources are used that can't emit infinitely squeezed light, the nullifiers are known to will not be equal to zero. Therefore to verify if the state is a cluster, not nullifiers themselves should be utilized, but rather their combination as given in the separability criteria of van Loock-Furusawa \cite{Furusawa}. In the case of $m$'th two-node cluster state this criterion provides the following condition for inseparability (entanglement) of two nodes:
\begin{multline} \label{7}
 \langle \left(\delta \hat{N}_m^1\right)^2 \rangle+\langle \left(\delta \hat{N}_m^2\right)^2 \rangle  \\
 = \langle \left(\delta \hat{Y}_m^1-\delta \hat{X}_m^2 \right)^2 \rangle+\langle \left(\delta \hat{Y}_m^2-\delta \hat{X}_m^1 \right)^2  \rangle
\textless \frac{1}{2},
\end{multline}
where $ \hat{N}_m^1$ and  $\hat{N}_m^2$ are nullifiers of first and second nodes of the $m$'th cluster state, respectively. Substituting in this criterion Fourier transforms of quadratures of the fields obtained after passing the scheme shown in Fig. \ref{Fig_1}, we can rewrite Eq. (\ref{7}) via $\hat{y}$ - quadratures of initially squeezed laser pulses (see Appendix A):
\begin{align}
2 \left(\langle |\delta \hat{y}^1_m(\omega _k)|^2 \rangle+ \langle |\delta \hat{y}^2_m (\omega _k)|^2 \rangle \right) \textless \frac{1}{2},
\end{align}
where $\hat{y}^1_m (\omega_k)$ and $\hat{y}^2_m (\omega _k)$ are Fourier transforms of squeezed $\hat{y}$-quadratures of phase-locked sub-Poissonian laser pulses, $\omega_k =2\pi k/(n(T+T_0))$ at $k \in \mathbf{Z}$ and $n \in \mathbf{N}$ is discrete  frequency, the introduction of which will be explained below in the section \ref{4A}. Since the lasers used have the equal variances, the condition of the inseparability of cluster state can be rewritten as:
\begin{align} \label{8}
4 \langle |\delta \hat{y}^1_m(\omega _k)|^2 \rangle \textless \frac{1}{2}.
\end{align}
Therefore, it follows that the cluster state inseparability depends on the variances of squeezed $\hat{y}$-quadratures by which this state was obtained. Substituting in the Eq. (\ref{8}) the variances of squeezed $ \hat{y} $-quadrature of the pulses of phase-locked sub-Poissonian laser  \cite{Korolev_2017}, we obtain an explicit dependence:
\begin{align} \label{12}
4\langle |\delta \hat{y}^1_m(\omega _k)|^2 \rangle=\frac{\omega _k ^2}{\kappa ^2+ \omega _k ^2} \textless \frac{1}{2}.
\end{align}
As this condition demonstrates, the inseparability of cluster state holds for frequencies $\omega_k \in \left(-\kappa ,\kappa \right)$. Moreover, for frequency $\omega_k =0$ the right-hand side has its minimum value, which corresponds to the largest squeezing of the oscillator on this frequency. From the physical standpoint, such oscillators correspond to pulses as a whole, with amplitudes
\begin{align}\label{field}
\hat{E}_m^j \equiv \hat{E}_m^j (0)=\hat{Q}_m^j(0)+i\hat{P}_m^j(0) =\int \limits _{-\infty}^{+\infty} \hat{E}_m^j(t)\;dt .
\end{align}
To measure such pulses the procedure of homodyne detection is to be used: signals under study are mixed with the local oscillators, the time profiles of which fully match profiles of measured pulses, and detectors average incoming signals by time. It can be concluded, that cluster states will be the best if for their generation pulses will be used as a whole. Exactly this case will be considered further.

\section{Computations on two-node cluster state}
It is possible now to begin the discussion of the quantum Gaussian computations on two-node cluster states. The main objective of any quantum computation is the transformation of some quantum states (taken as input) into different ones (output). Moreover, in order for the computing model to be considered universal, it should be able to carry out any kind of transformations. As shown in \cite{Gu}, a quantum multimode Gaussian transformation can be represented as a result of the sequential application of two types of transformations: arbitrary single-mode transformations and one controlled two-mode transformation. Thus, to confirm the implementation of universal quantum calculations on the ensembles of two-node cluster states, both types of presented operations need to be demonstrated.

In this section, as a starting point, the implementation of the single-mode operation is described. Initially, the computation procedure is discussed formally and then its specific example of implementation is demonstrated.

Using a symmetric beam splitter, the input state $ \hat{A} _ {in} (t) = \hat {x} _ {in} (t) + i \hat {y} _ {in}(t) $ to be transformed is mixed with the upper node of the cluster state obtained as described in the previous section. Field amplitudes of the input and cluster states after such transformation may be written as:
\begin{multline}
\hat{A}_{in}(t) \rightarrow \hat{A}'_{in}(t)=\frac{1}{\sqrt{2}}\left(\hat{x}_{in}(t)+\hat{x}_m^1(t) +\hat{y}^2_m(t)\right)\\
+\frac{i}{\sqrt{2}}\left( \hat{y}_{in}(t)- \hat{x}_m^2(t)+\hat{y}_m^1(t)\right),
\end{multline}
\begin{multline}
\hat{E}_{m}^1(t) \rightarrow\hat{E}_m^{1\prime}(t)=\frac{1}{\sqrt{2}}\left(\hat{x}_{in}(t)-\hat{x}_m^1(t) -\hat{y}^2_m(t)\right)\\
+\frac{i}{\sqrt{2}}\left( \hat{y}_{in}(t)+ \hat{x}_m^2(t)-\hat{y}_m^1(t)\right),
\end{multline}
\begin{multline}
\hat{E}_m^2(t)=-\frac{1}{\sqrt{2}}\left(\hat{x}_m^2 (t)+\hat{y}^1_m(t)\right)\\
+\frac{i}{\sqrt{2}}\left( \hat{x}_m^1(t)-\hat{y}_m^2(t)\right),
\end{multline}
where $\hat{x}_m^j(t)$ and $\hat{y}_m^j(t)$ are the quadratures of the $m$'th pulse laser field in the $j$'th channel, $\hat{E}_m^j(t)$ are the amplitudes of cluster states in the $j$'th channel. Next, two light pulses at the output of the beam splitter are measured by balanced homodyne detectors (pulses with amplitudes $ \hat{A}'_ {in} (t) $ and $ \hat {E} _m ^{1 \prime} (t) $).  The measurements of these two pulses affect the unmeasured pulse $\hat{E}_m^2(t)=\hat{X}_m^{2}(t)+i\hat{Y}_m^{2}(t)$, since as was shown earlier the cluster state is inseparable (entangled). In general, this effect can be written as \cite{Ukai}:
\begin{multline} \label{13}
\begin{pmatrix}
\hat{X}_m^{2}(t)\\
\hat{Y}_m^{2}(t)
\end{pmatrix} =\mathcal{L}(t)\Big[M(\theta_+,\theta_-)\begin{pmatrix}
\hat{x}_{in}\\
\hat{y}_{in}
\end{pmatrix}-\sqrt{2}\begin{pmatrix}
\hat{y}_{m}^1\\
\hat{y}_m^{2}
\end{pmatrix}  \\
+\frac{1}{\beta _0\sqrt{2}\sin \theta _{-}}\begin{pmatrix}
\cos \theta _{1} & -\cos \theta _{in}\\
-\sin \theta _1 & \sin \theta _{in}
\end{pmatrix} \begin{pmatrix}
{i}_{in}\\
{i}_{1}
\end{pmatrix}\Big],
\end{multline}
where $\theta _{in}$, $\theta _1$ are phases, and $\beta _0$ is an amplitude of local oscillators, used for homodyne detection of light pulses, $\mathcal{L}(t)$ is the time envelope of the detected signal, $\theta _{\pm}=\theta _{in} \pm \theta _1$, and the matrix $M(\theta_+,\theta_-)$ has the following form:
\begin{align}  \label{M_1}
M(\theta_+,\theta_-)=\frac{1}{\sin \theta _{-}}\begin{pmatrix}
\cos \theta _{+}+\cos \theta _{-} & \sin \theta _{+}\\
-\sin \theta _{+} & \cos \theta _{+}-\cos \theta _{-}
\end{pmatrix}.
\end{align}
To derive the Eq. (\ref{13}) it was taken into account that the homodyne detections are carried out with the local oscillator, the time profile of which exactly matches the profile of the measured pulse, and detectors average the incoming signal over time. This derivation is discussed in more detail in \cite{me2}.

Let us consider the structure of the derived Eq. (\ref{13}) in more detail. It consists of three terms. The last term consists only of classical values and, therefore, its impact on quadratures can be fully compensated. The second term includes only squeezed quadratures of the laser radiation, which can be made small given sufficient enough squeezing (this component is responsible for the calculation error). The first term is the transformation itself that is performed as a result of calculations on quadratures $\hat{x}_{in}$ and $\hat{y}_{in}$. Let us note that this transformation depends only on the phases of local oscillators of homodyne detectors. By varying these phases, the specific transformation of the input state can be chosen. Nevertheless, the executed transformation is not yet universal.

If the field obtained in these calculations is sent to the input of another identical transformation, the resulted quadratures can be written as:
\begin{align} \label{14}
\begin{pmatrix}
\hat{X}_{out}(t)\\
\hat{Y}_{out}(t)
\end{pmatrix} =\mathcal{L}(t)M(\theta_{+}^\prime,\theta_{-}^\prime)\;M(\theta_{+},\theta_{-})\begin{pmatrix}
\hat{x}_{in}\\
\hat{y}_{in}
\end{pmatrix},
\end{align}
where $\theta _{\pm}^\prime$ is determined by the phases of the local oscillators of the second transformation. It was shown \cite{Ukai}, that Eq. (\ref{14}) is the single-mode universal Gaussian transformation. Hence, it can be concluded, that the universal single-mode transformation requires an ensemble of a pair of two-node cluster states.

The implementation of such transformation is described below, based on the calculation scheme presented in \cite {OPO}.
\begin{figure*}
\centering
\includegraphics[scale=0.95]{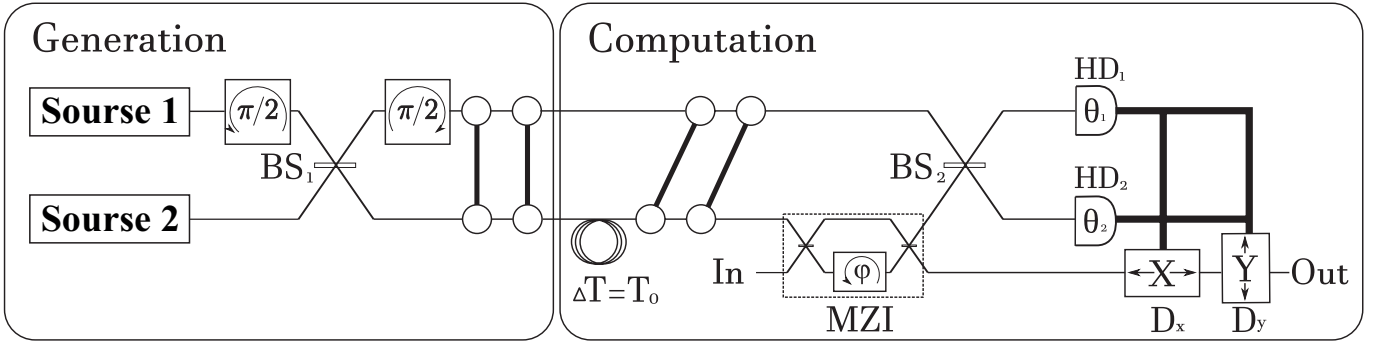}
\caption{Scheme of the one-way computations with an ensemble of two-node cluster states. The scheme includes two parts: the generation of the cluster
ensemble and the calculation itself. On the figure \texttt{Source 1} and \texttt{Source 2} are pulse sources with the squeezed $\hat{y}$-quadrature,
which consist of synchronized phase-locked lasers, phase shifters and apertures (see Fig. \ref{Fig_1}). \texttt{BS$_1$} and \texttt{BS$_2$} are beam
splitters, \texttt{$\Delta T=T_0 $} is a time delay, \texttt{In} is a channel used by the Mach-Zehnder interferometer (\texttt{MZI}) to input unknown
state for further calculations. \texttt{{HD}$_1$} and \texttt{{HD}$_2$} are two balanced homodyne detectors. Results obtained by detectors are sent
to units that shift quadratures by the required classic value. \texttt{Out} is the channel that outputs state of interest after all calculations.}
\label{Fig_2}
\end{figure*}
Let us consider the implementation of the single-mode computation which is carried out on the ensemble of four squeezed pulses from the phase-locked sub-Poissonian lasers (see Fig.\ref{Fig_2}). The first part of the scheme transforms these pulses into a pair of two-node cluster states separated by the time interval $T_0$. These states are fed into the second part responsible for computation. The first element of this part is the delay line that shifts all pulses in the lower arm by the time $T_0$. Next, the Mach-Zehnder interferometer with the phase shifter in its lower arm is placed in the scheme. Due to phase shifting by $\varphi =\pi $ and $\varphi =0$, this interferometer can serve as a switcher. The interferometer's scheme of operation and transformation steps are detailed in Fig. \ref{Fig_3}.
\begin{figure*}
\centering
\includegraphics[scale=0.95]{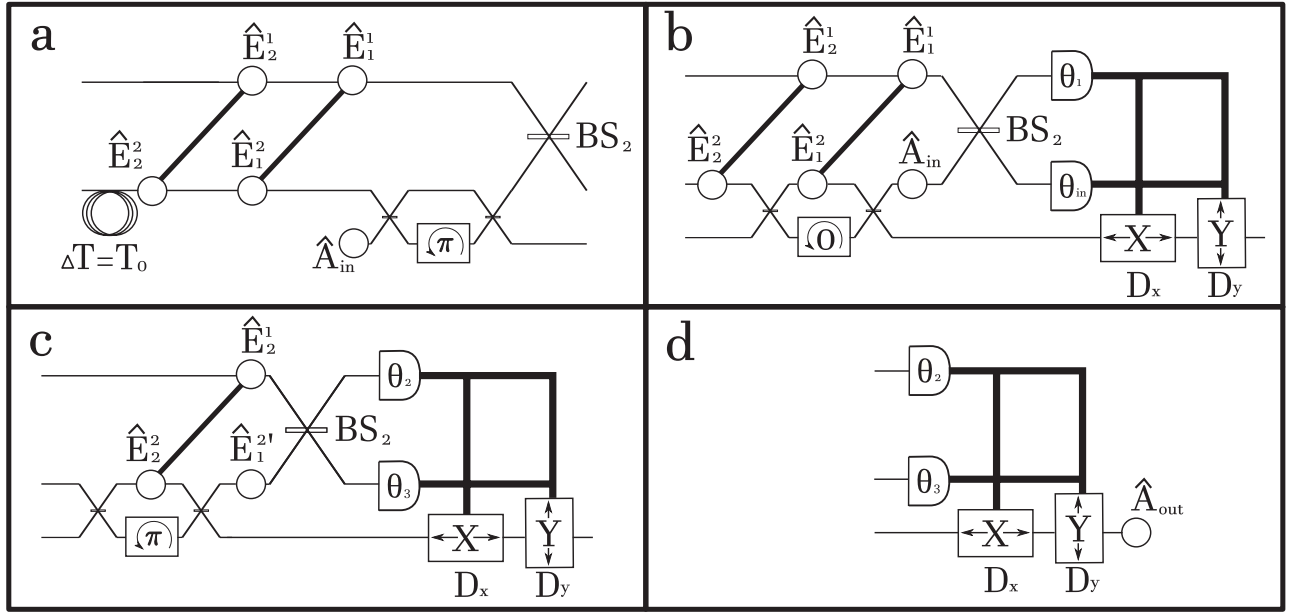}
\caption{Description of each stage of the calculation is depicted in Fig. \ref{Fig_2}. } \label{Fig_3}
\end{figure*}
At the first stage, after setting the value $\varphi=\pi$, the state $\hat{A}_{in}$ is inputted to the calculation in such a way as to assure that it is synchronized with the state $\hat E_1^1$ in the upper arm by time (Fig. \ref{Fig_3}a). After this the interferometer's phase $\varphi =0$ is set. Then the pulses getting into the upper arm will leave also from that arm (Fig. \ref{Fig_3}b,c). At the same time, the first pulse of the cluster state in the upper arm and the input pulse in the lower arm are fed into the second beam splitter $\mbox{BS}_2$, where they are mixed and then measured by two homodyne detectors $\text {HD}_1$ and $\text {HD}_2$. These measurements, according to Eq. (\ref{13}), affect the quadratures of the field $\hat{E}^2_1$ (on Fig. \ref{Fig_3} the modified field is denoted by the operator $\hat E_1^{2\prime}$). This field is passed through the lower arm and coincides in time with $\hat{E}_2^1$, hence this is the input field for calculations performed on the next cluster state. The quadratures of the field $\hat{E}_2^2$ have changed according to the Eq. (\ref{14}) after  $\hat{E}_2^1$ and $\hat{E}_1^{2\prime}$ pass through the beam splitter $\text{BS}_2$  and are measured. At the same time, the interferometer's phase is set to $\varphi =\pi$ again, and the state arising after all transformations passes through the interferometer leaving it in the lower arm. After that this state arrives at displacement devices ($\text{D}_x$ and $\text{D}_y$), which take into account the results of previous measurements and shift the quadratures in such a way that no classical values are left in them (Fig. \ref{Fig_3}d). Thereby, using the scheme depicted in. Fig. \ref{Fig_2}, the universal single-mode transformations over the input state $\hat{A}_{in}$ can be carried out.

\subsection{ Parallel computations} \label{4A}
\begin{figure*}
\centering
\includegraphics[scale=1.2]{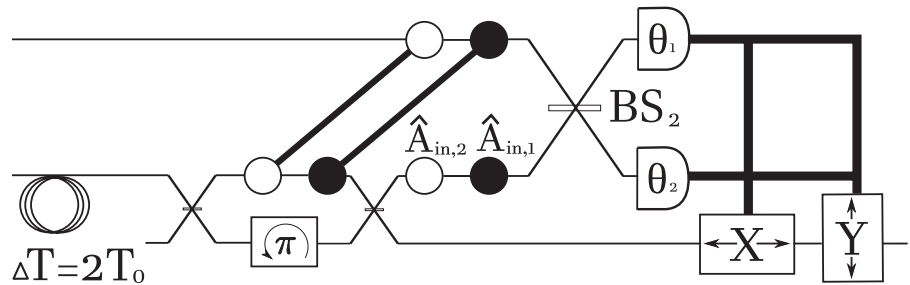}
\caption{Scheme of parallel computations on the two input states $\hat{A}_{in,1}$ and $\hat{A}_{in,2}$. Black and white circles mark the modes, which
do not interact with each other at computations.} \label{Fig_4}
\end{figure*}

The significant weakness of the scheme described earlier is the fact that it can transform an one input state. If there are several input states, the transformations of each one can be performed only after the computation of previous states has been finalized. However, this scheme can be improved. For this purpose, the delay in the interferometer's lower arm needs to be increased. For example, let us consider the delay in $2\text{T}_0$. Here, as before, it is assumed that there are four squeezed pulses from sub-Poissonian lasers present, from which the pair of two-node cluster states is generated. Let us consider how this scheme can be used for the parallel computation on two input states (see the explanatory Fig. \ref{Fig_4}).
\\
Assume that two states $\hat{A}_{in,1}(t)$, $\hat{A}_{in,2}(t)$ are got into the input with the time interval $T_0$. Due to the absence of the links between two cluster states during transformations, first modes, marked by black on the figure, don't affect the second modes, marked by white. This results in completely parallel calculations on two input states. Obviously, the increase in delay also increases the number of modes in which parallel calculations can be carried out.

Let us consider now how many input states it is possible to enter into the scheme. In other words,  it needs to be determined on which time entangled pulses can be delayed while conserving the entanglement. To answer this, the van Loock-Furusawa criterion for the cluster state is analyzed in which one of the pulses is delayed by the time $\tau =n(T+T_0)$ at any $n \in \mathbf{N}$. Amplitudes of such pulses in time representation can be written as:
\begin{multline}
\hat{E}_m^1(t)=\frac{1}{\sqrt{2}}\left(\hat{x}_m^1(t) +\hat{y}^2_m(t)\right)-\frac{i}{\sqrt{2}}\left( \hat{x}_m^2(t)-\hat{y}_m^1(t)\right),
\end{multline}
\begin{multline}
\hat{E}_m^2(t-\tau)=-\frac{1}{\sqrt{2}}\left(\hat{x}_m^2(t-\tau) +\hat{y}^1_m(t-\tau)\right)\\
+\frac{i}{\sqrt{2}}\left(
\hat{x}_m^1(t-\tau)-\hat{y}_m^2(t-\tau)\right).
\end{multline}
Let us continue with the Fourier transform of these values and define the canonical quadratures $\hat{Q}_m^j(\omega_k)$ and $\hat{P}_m^j(\omega_k)$ using the frequency sampling.
\begin{multline}
\hat{Q}_m^1(\omega_k)=\frac{1}{\sqrt{2}}\Big( \Re \left[\hat{x}_m^1(\omega_k)+ \hat{y}_m^2(\omega_k)\right]  \\
- \Im \left[ \hat{y}_m^1(\omega_k)-\hat{y}_m^2(\omega_k)\right]\Big),
\end{multline}
\begin{multline}
\hat{P}_m^1(\omega_k)=\frac{1}{\sqrt{2}}\Big( \Im \left[\hat{x}_m^1(\omega_k)+ \hat{y}_m^2(\omega_k)\right] \\
+\Re \left[ \hat{y}_m^1(\omega_k)-\hat{y}_m^2(\omega_k)\right]\Big),
\end{multline}
\begin{multline}
\hat{Q}_m^2(\omega_k)=-\frac{1}{\sqrt{2}}\Big(\Re \left[e^{\displaystyle i\omega_k \tau }\left(\hat{x}_m^2(\omega_k)+ \hat{y}_m^1(\omega_k)\right)\right] \\
-\Im \left[e^{\displaystyle i\omega_k \tau }\left(\hat{y}_m^2(\omega_k)- \hat{x}_m^1(\omega_k)\right)\right]\Big),
\end{multline}
\begin{multline}
\hat{P}_m^2(\omega_k)=-\frac{1}{\sqrt{2}}\Big(\Im \left[e^{\displaystyle i\omega_k \tau }\left(\hat{x}_m^1(\omega_k)+
\hat{y}_m^2(\omega_k)\right)\right] \\
+\Re \left[e^{\displaystyle i\omega_k \tau }\left(\hat{y}_m^2(\omega_k)-
\hat{x}_m^1(\omega_k)\right)\right]\Big).
\end{multline}
Substituting the derived quadratures into the van Loock-Furusawa criterion for the two-node cluster state (\ref{7}), we obtain
\begin{multline}
2\cos ^2\left(\frac{\omega _k\tau}{2}\right)\left(\langle |\delta \hat{y}_m^1(\omega_k)|^2\rangle +\langle |\delta
\hat{y}_m^2(\omega_k)|^2\rangle\right)\\
+2\sin ^2\left(\frac{\omega_k \tau}{2}\right)\left(\langle |\delta \hat{x}_m^1(\omega_k)|^2\rangle +\langle
|\delta \hat{x}_m^2(\omega_k)|^2\rangle\right) \textless \frac{1}{2},
\end{multline}
where $\hat{x}_m^j(\omega_k)$ is the Fourier transform of the stretched $\hat{x}$-quadrature of sub-Poissonian laser pulse. Since the used lasers radiate the light with the equal variances, so the separability criterion can be rewritten in the form:
\begin{multline} \label{vanLook}
4\cos ^2\left(\frac{\omega _k\tau}{2}\right)\langle |\delta \hat{y}_m^1(\omega_k)|^2\rangle\\
+4\sin ^2\left(\frac{\omega_k \tau}{2}\right)\langle
|\delta \hat{x}_m^1(\omega_k)|^2\rangle \textless \frac{1}{2}.
\end{multline}
Since for the utilized states  $\langle |\delta \hat{x}_m^1(\omega_k)|^2\rangle \gg \langle |\delta \hat{y}_m^1(\omega_k)|^2\rangle$, then the inequality (\ref{vanLook}) is true for discrete frequencies $\omega_k=2\pi k /\tau \equiv 2\pi k /n(T+T_0)$ for any $k \in \mathbf{Z}$ and $n \in \mathbf{N}$. Thus, at the right sampling frequency the van Loock-Furusawa criterion is fulfilled for any delay intervals that are multiple of the pulse period. Moreover, for these frequencies the criterion reduces to the condition (\ref{12}), which is performed best for oscillators at zero frequency. Summing up, it can be concluded that the best quantum entanglement of the cluster state is obtained using the pulses as a whole (oscillators at zero frequencies) and that one of the pulses of this state can be delayed for any duration as long as it is multiple of the pulse period. It should be noted, though, that this conclusion does not take into account dissipative losses usually present in real physical channels that limit the lifetime of entangled states.

\subsection{Two-mode operations}
It appears that using parallel calculations not only allows us to perform the independent universal single-mode calculations over the different input
modes but also the two-mode transformations needed for the realization of the universal quantum computer. Let us consider the
$\text{CZ}$-transformation as an example of a two-mode one. By definition, this transformation turns the quadratures of the input fields into the
output quadratures by the following rule
\begin{align} \label{CZ}
\begin{pmatrix}
\hat{X}_{out,1}\\
\hat{Y}_{out,1}\\
\hat{X}_{out,2}\\
\hat{Y}_{out,2}
\end{pmatrix}=\begin{pmatrix}
1 & 0 & 0 & 0\\
0 & 1 & 1 & 0\\
0 & 0 & 1 & 0 \\
1 & 0 & 0 & 1
\end{pmatrix}\begin{pmatrix}
\hat{x}_{in,1}\\
\hat{y}_{in,1}\\
\hat{x}_{in,2}\\
\hat{y}_{in,2}
\end{pmatrix}.
\end{align}

\begin{figure*}
\centering
\includegraphics[scale=1]{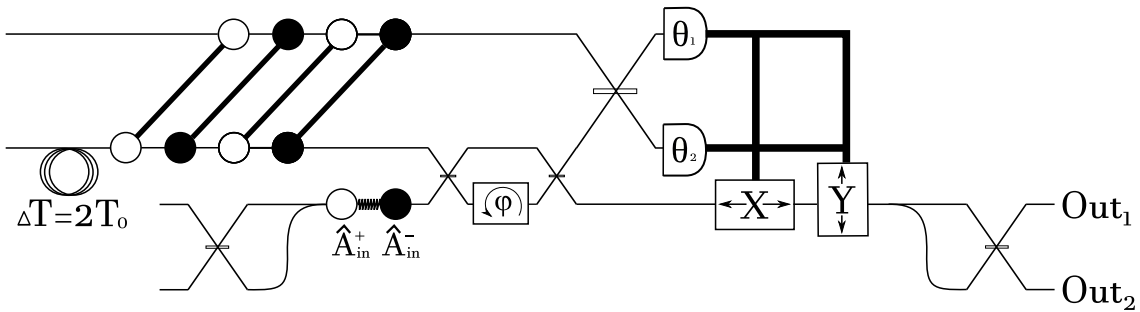}
\caption{Scheme of CZ operation with two sub-Poissonian lasers. Here $\Delta T =2T_0$ is the pulse delay in the interferometer's lower arm, $\hat{A}_{in}^{\pm}= \frac{1}{\sqrt{2}}(\hat{A}_{in,1}\pm\hat{A}_{in,2})$ are superpositions of the input fields after passing through the beam splitter, $\mbox{Out}_1$ and $\mbox{Out}_2$ are output channels for the computation results.} \label{Fig_5}
\end{figure*}

To implement this transformation, the scheme presented in Fig. \ref{Fig_4} needs to be modified slightly. Namely, two additional beam splitters have to be placed at the scheme’s input and output. The resulting scheme is depicted in Fig. \ref{Fig_5}.
In this scheme, the pulses are delayed in the lower arm by the time $2T_0$, which sets up the scheme for the parallel transformation of two states. The superpositions $\hat{A}_{in}^{+}(t-T-T_0)=1/\sqrt{2}(\hat{A}_{in,1}(t-T-T_0)+\hat{A}_{in,2}(t-T-T_0))$ and $\hat{A}_{in}^{-}(t)=1/\sqrt{2}(\hat{A}_{in,1}(t)-\hat{A}_{in,2}(t))$ are used as these states, which are obtained after states $\hat{A}_{in,1}(t)$ and $\hat{A}_{in,2}(t)$ have passed through the beam splitter and the corresponding delay of one of states. Two output states from the transformation scheme are passed through the identical beam splitter. The result of these operations can be written in matrix form as
\begin{widetext}
\begin{align}
\begin{pmatrix}
\hat{X}_{out,1}(t)\\
\hat{Y}_{out,1}(t)\\
\hat{X}_{out,2}(t)\\
\hat{Y}_{out,2}(t)
\end{pmatrix}=\mathcal{L}(t)\begin{pmatrix}
\frac{1}{\sqrt{2}} & 0 & \frac{1}{\sqrt{2}} & 0\\
0 & \frac{1}{\sqrt{2}} & 0 & \frac{1}{\sqrt{2}}\\
\frac{1}{\sqrt{2}} & 0 & -\frac{1}{\sqrt{2}} & 0 \\
0 & \frac{1}{\sqrt{2}} & 0 & -\frac{1}{\sqrt{2}}
\end{pmatrix}\begin{pmatrix}
a_{11} & a_{12} & 0 & 0\\
a_{21} & a_{22} & 0 & 0\\
0 & 0 & b_{11} & b_{12} \\
0 & 0 & b_{21} & b_{22}
\end{pmatrix}\begin{pmatrix}
\frac{1}{\sqrt{2}} & 0 & \frac{1}{\sqrt{2}} & 0\\
0 & \frac{1}{\sqrt{2}} & 0 & \frac{1}{\sqrt{2}}\\
\frac{1}{\sqrt{2}} & 0 & -\frac{1}{\sqrt{2}} & 0 \\
0 & \frac{1}{\sqrt{2}} & 0 & -\frac{1}{\sqrt{2}}
\end{pmatrix}\begin{pmatrix}
\hat{x}_{in,1}\\
\hat{y}_{in,1}\\
\hat{x}_{in,2}\\
\hat{y}_{in,2}
\end{pmatrix},
\end{align}
\end{widetext}
where $a_{ij}$ and $b_{ij}$ are matrix elements of two single-mode transformations. If one choose linear optics elements corresponding to $a_{11}=a_{22}=a_{21}=b_{11}=b_{22}=1$, $a_{12}=b_{12}=0$, $b_{21}=-1$, the transformation matches (\ref{CZ}) to an accuracy of the classical time amplitude $\mathcal{L}(t)$, which point at the profile of signal whose quadratures are transformed. Thus, using two additional beam splitters and the parallel computing scheme depicted in Fig. \ref{Fig_4}, two-mode operations over input states can be carried out.

\section{Conclusion}
This study shows that using the ensemble of two-node cluster states and the parallel computation scheme presented in Fig. \ref{Fig_4}, the arbitrary single-mode Gaussian computations over a large number of states as well as two-mode operations can be carried out. As previously noted, this set of operations ensures the implementation of universal Gaussian transformations on input modes. Moreover, the study \cite{Gu} demonstrates the implementation of non-Gaussian operations on two-node cluster states. This is sufficient enough to suggest the possibility of universal quantum computations on the two-node cluster state ensemble. To realize the quantum computer, the capability to correct errors arising in the calculation process has to be provided. In one-way computations, the cluster states are usually considered for this purpose which has the supportive nodes not used in the calculations themselves but allowing to perform the procedure of the error correction \cite{Nielsen}. In our case, the minimal possible squeezed state is considered which does not allow the additional connections. Consequently, the problem of the error correction in our case requires further analysis. For example, the error correction procedure associated with limited squeezing, proposed in  \cite{Su}.

It was demonstrated how the logical gates can be performed using the sequence of pulses from two synchronized sub-Poissonian lasers. Operating with
this light source has some known advantages with possible experimental implementation. The maximum squeezing in sub-Poissonian lasers is reached high
above the lasing threshold, in the saturation regime. In OPO, for example, on the contrary, squeezed light can be obtained only near the threshold.
Since the near-threshold region of generation is not stable for any laser system, the sub-Poissonian laser provides an easier and more reliable
generation of squeezed light. Operating high above the lasing threshold is also more convenient with regard to the homodyne detection: the space-time
profile of the laser radiation can be used as a homodyne, which greatly simplifies the experiment. Moreover, sub-Poissonian light sources are a more
readily available alternative to parametrical systems. It is no mere chance that the first commercial quantum cryptographic devices \cite{IDQ, MagiQ}
were based on the radiation of this kind of lasers. It should also be noted that the excess noise arising in the stretched quadrature of the laser
radiation does not affect the efficiency of the scheme \cite{Korolev_2017,Walshe}.

Experimental schemes for continuous-variable quantum computing are mostly oriented on creating large branched cluster structures
\cite{Menicucci_2011,Ukai}. From this point of view, our proposal seems more promising, since it does not require the generation and preservation of
a large-scale entangled state. Estimating the possibilities of experimental implementation of quantum computations on cluster states, we can identify
two main sources of errors that may prevent successful computations. The first type of errors is associated with the possible nonideality of optical
elements, which are used to cluster states generation and the subsequent quantum computation. Our schemes minimize such errors by minimizing the
number of optical elements in them. In addition, the optical elements used in our schemes are easily implemented in practice. For example, one of the
problems of the experimental realization of cluster states is the need to create beam splitters with very asymmetrical coefficients of transmission
and reflection \cite{Ukai}. Alternatively, there are only symmetric beam splitters in this proposal. The second type of error obstructed the quantum
computation is associated with the non-ideal squeezing of the light used. It has been experimentally demonstrated that the lasers with regular
pumping can generate light with squeezing of $8.3$ dB \cite{Yamamoto}. Although this value is significantly less than the theoretical threshold value
15 -- 17 dB, which provides fault-tolerance in the execution of arbitrary computational protocols \cite{Walshe}, such a resource allows to generate
two-node cluster states and carry out a significant number of computations on them.

\section*{Acknowledgement}
This work was supported by the Russian Foundation for Basic  Research  (Grant  Nos 18-32-00255mol\_a, 19-02-00204a, and  18-02-00648a ). Scientific results were achieved during the implementation of the Program within the state support of the STI Center "Quantum Technologies".

\appendix
\section{Van Loock-Furusawa criterion for two-node cluster states, obtained by sub-Poissonian lasers}
Let us write the van Loock-Furusawa criterion for two-node cluster states:
\begin{multline} \label{a1_1}
 \langle \left(\delta \hat{N}_m^1\right)^2 \rangle+\langle \left(\delta \hat{N}_m^2\right)^2 \rangle  \\
= \langle \left(\delta \hat{Y}_m^1-\delta \hat{X}_m^2 \right)^2 \rangle+\langle \left(\delta \hat{Y}_m^2-\delta \hat{X}_m^1 \right)^2  \rangle
\textless \frac{1}{2}.
\end{multline}
Let us check the fulfilment of this criterion for the cluster state obtained with the pulse radiation of sub-Poissonian synchronized lasers. For this purpose, the quadratures of the state obtained after passing the depicted in Fig. \ref{Fig_1} scheme should  be substituted into nullifiers $\hat{N}_m^1$ and $\hat{N}_m^2$. These quadratures can be rewritten through the quadratures of initially squeezed oscillators in the form:
\begin{multline} \label{a1_2}
\hat{E}_m^1(t)=-\frac{i}{\sqrt{2}}\left(i\hat{S}_m^1(t)+\hat{S}_m^2(t)\right) \\
=\frac{1}{\sqrt{2}}\left(\hat{x}_m^1(t) +\hat{y}^2_m(t)\right)-\frac{i}{\sqrt{2}}\left( \hat{x}_m^2(t)-\hat{y}_m^1(t)\right)  \\
\equiv\hat{X}^1_m(t)+i\hat{Y}_m^1(t),
\end{multline}
\begin{multline}\label{a1_3}
\hat{E}_m^2(t)=\frac{1}{\sqrt{2}}\left(i\hat{S}_m^1(t)-\hat{S}_m^2(t)\right)  \\
=-\frac{1}{\sqrt{2}}\left(\hat{x}_m^2(t)
+\hat{y}^1_m(t)\right)+\frac{i}{\sqrt{2}}\left( \hat{x}_m^1(t)-\hat{y}_m^2(t)\right) \\
\equiv \hat{X}^2_m(t)+i\hat{Y}_m^2(t).
\end{multline}

The mathematical procedure to identify the canonical variables needs to be discussed before the substitution of the resulting amplitudes to Eq. (\ref{a1_1}) for nullifiers. As was said earlier, the canonical quadratures which commute on the number are used in the equation for nullifiers. For employed sources, the quadratures commute on Dirac delta-function. To correct this situation the following Fourier-transformation is applied to the slow amplitude envelope of laser fields
\begin{align}
\hat{E}^j_m(\omega)=\int \limits _{-\infty}^{\infty} dt \;\hat{E}^j_m(t)e^{\displaystyle i \omega t}.
\end{align}
As the result, the commutation relations can be written as:
\begin{multline}
\left[\hat{E}_m^j(\omega),\left(\hat{E}_{m'}^{ j'} ({ \omega}')\right)^{\dag} \right]\\
=\iint _{-\infty}^{\infty} \left[
\hat{E}_m^j(t),\left(\hat{E}_{m'}^{j'}({t}')\right)^{\dag} \right] e^{\displaystyle i \omega t} e^{\displaystyle i {\omega}' {t'}} dt dt' \\
=\delta
_{jj'} \delta _{mm'}\delta (\omega+{\omega}').
\end{multline}
If we pass to the discrete frequencies $\omega_k=2\pi k/n(T+T_0)$ at $k \in \mathbf{Z}$ and $n\in \mathbf{N}$, the final commutation relation can be given as
\begin{multline}
\left[\hat{E}_m^j(\omega_k),\left(\hat{E}_{m'}^{j'}(\omega _{k'})\right)^{\dag} \right]=\delta _{jj'}\delta _{m m'}\delta _{\omega _k,-\omega _{k'}},\\
\Rightarrow \left[\hat{Q}_m^j(-\omega_k),\hat{P}_m^{j} (\omega _k) \right]=\frac{1}{4},
\end{multline}
where $\hat{Q}_m^j(\omega _k)$ and $\hat{P}_m^j(\omega _k)$ are the quadratures of $m$'th light pulse in $j$-channel on the discrete frequency $\omega
_k$, connected with the Fourier transform of time quadratures by relationships
\begin{align} \label{a1_4}
&\hat{Q}_m^j(\omega_k)=\Re \hat{E}_m^j (\omega_k) =\Re  \hat X_m^j (\omega_k)-\Im \hat{Y}_m^j (\omega_k),\\
&\hat{P}_m^j(\omega_k)=\Im \hat{E}_m^j (\omega_k) =\Im  \hat X_m^j (\omega_k)+\Re \hat{Y}_m^j (\omega_k). \label{a1_5}
\end{align}
These quadratures are canonical and that is why they have to be used in the van Loock-Furusawa criterion (\ref{a1_1}):
\begin{multline}
\langle \left(\delta \hat{P}_m^1(\omega_k)-\delta\hat{Q}_m^2 (-\omega_k)\right)^2 \rangle \\
+\langle \left(\delta
\hat{P}_m^2(\omega_k)-\delta\hat{Q}_m^1 (-\omega_k)\right)^2 \rangle \textless \frac{1}{4}.
\end{multline}
Substituting here the relations (\ref{a1_2}), (\ref{a1_3}) and taking into account (\ref{a1_4}) and (\ref{a1_5}), the van Loock-Furusawa criterion for the two-node cluster state can be expressed in terms of initially squeezed quadratures of sub-Poissonian laser:
\begin{align}
2 \left(\langle |\delta \hat{y}^1_m(\omega _k)|^2 \rangle+ \langle |\delta \hat{y}^2_m (\omega _k)|^2 \rangle \right) \textless \frac{1}{2}.
\end{align}


\begin{thebibliography}{100}
\bibitem{Menicucci_1}  N. Menicucci et. al., Phys. Rev. Lett. \textbf{97}, 110501 (2006).
\bibitem{Raus2} R. Raussendorf and H. J. Briegel, Phys. Rev. Lett. \textbf{86}, 5188 (2001).
\bibitem{Korolev_2018} S. B. Korolev et al., Laser Phys. Lett. \textbf{15}, 075203 (2018).
\bibitem{Lloyd} S. Lloyd, S. L. Braunstein, Phys. Rev. Lett. \textbf{82}, 1784 (1999).
\bibitem{Menicucci_2011} N. C. Menicucci,  Phys. Rev. A  \textbf{83}, 062314 (2011).
\bibitem{Gu} M. Gu et. al., Phys. Rev. A \textbf{79}, 062318 (2009).
\bibitem{Marek} P. Marek, R. Filip, and A. Furusawa, Phys. Rev. A \textbf{84}, 053802 (2011).
\bibitem{Yukawa} M. Yukawa et. al., Phys. Rev. A \textbf{88}, 053816 (2013).
\bibitem{Park} K. Park, P.Marek, and R. Filip, Phys. Rev. A \textbf{90}, 013804 (2014).
\bibitem{Arzani} F. Arzani, N. Treps, G. Ferrini, Phys. Rev. A \textbf{95}, 052352 (2017).
\bibitem{Kim} J. Kim, S. Sonami, and  Y. Yamamoto, Nonclassical Light from
Semiconductor Lasers and LEDs (Springer, Berlin, 2001).
\bibitem{Yamamoto} W. H. Richardson , S. Machida, Y. Yamamoto, Phys. Rev. Lett. \textbf{66}, 2867 (1991)
\bibitem{HAL} G. Ferrini et al., Phys. Rev. A  \textbf{91}, 032314 (2015).
\bibitem{Korolev_2017} S.B. Korolev , K. S. Tikhonov , T. Yu. Golubeva, and Yu. M. Golubev, Optics and Spectroscopy  \textbf{123} (3), 411 (2017).
\bibitem{vLoock} P. van Loock, C. Weedbrook, and M. Gu, Phys. Rev. A \textbf{76}, 032321 (2007).
\bibitem{Menicucci} N. Menicucci, Phys. Rev. Lett. \textbf{112}, 120504 (2014).
\bibitem{Vahlbruch} H. Vahlbruch, M. Mehmet, K. Danzmann, R. Schnabel, Phys. Rev. Lett. \textbf{117}, 110801 (2016).
\bibitem{Golubeva}  T. Golubeva, D. Ivanov, and Yu. Golubev, Phys. Rev. A  \textbf{77}, 052316 (2008).
\bibitem{Samburskaya} K. S. Samburskaya, T. Yu. Golubeva, V. A. Averchenko, and Yu. M. Golubev,  Optics and Spectroscopy \textbf{113} (1), 86 (2012).
\bibitem{Furusawa} P. van Loock and  A. Furusawa, Phys. Rev. A  \textbf{67}, 052315 (2003).
\bibitem{Ukai} R. Ukai, Multi-Step Multi-Input One-Way Quantum Information Processing with Spatial and Temporal Modes of Light. (Tokyo: Springer Japan, 2015).
\bibitem{me2} S.B. Korolev, E.A. Vashukevich,  T.Yu. Golubeva, and  Yu.M. Golubev, Quantum Electronics \textbf{48} (10), 906 (2018).
\bibitem{OPO} S. Yokoyama  et al., Nat. Photon. \textbf{7}, 982 (2013).
\bibitem{Nielsen} M. A. Nielsen,  C. M. Dawson, Phys. Rev. A \textbf{71}, 042323 (2005).
\bibitem{IDQ} ID Quantique (IDQ), [Electronic resource] URL: http://www.idquantique.com
\bibitem{MagiQ} MagiQ, [Electronic resource] URL: http://magiqtech.com
\bibitem{Su} D. Su, C. Weedbrook, and K. Br\'adler, Phys. Rev. A \textbf{98}, 042304, (2018).
\bibitem{Walshe} B. W. Walshe, L. J. Mensen, B. Q. Baragiola, N. C. Menicucci,
 arXiv:1903.02162 [quant-ph] (2019)
\end{thebibliography}
\end{document}